\documentstyle[twoside,fleqn,espcrc2,epsf,epsfig]{article}
\def\lsim{\raise0.3ex\hbox{$<$\kern-0.75em\raise-1.1ex\hbox{$\sim$}}}
\def\gsim{\raise0.3ex\hbox{$>$\kern-0.75em\raise-1.1ex\hbox{$\sim$}}}
\def\be{\begin{equation}}
\def\ee{\end{equation}}
\def\ba{\begin{eqnarray}}
\def\ea{\end{eqnarray}}
\def\bea{\begin{eqnarray}}
\def\eea{\end{eqnarray}}
 2
\def\m0{m_{D0}}
\def\bx{{\bf x}}

\def\half{{1\over 2}}

\newcommand{\tr}{{\rm Tr}}
\def\d{\partial}

\begin{document}
\title{Screening in Hot SU(2) Gauge Theory and Propagators in
3d Adjoint Higgs model}
\author{
A. Cucchieri, F. Karsch and P. Petreczky\\
Fakult\"at f\"ur Physik, Universit\"at Bielefeld,
P.O. Box 100131, D-33501 Bielefeld, Germany
}
\begin{abstract}
We investigate the large distance behavior of the electric 
and magnetic propagators of hot SU(2) gauge theory in different gauges
using lattice simulations of the full 4d theory and the effective,
dimensionally reduced 3d theory. A comparison of the 3d
 and 4d data for 
the propagators suggests that dimensional reduction works surprisingly well
down to  temperatures $T \sim 2 T_c$. 
A detailed study of the volume dependence of magnetic propagators is 
performed.
The electric propagators show exponential decay at
large distances in all gauges considered and a possible gauge dependence 
of the electric screening mass
turns out to be statistically insignificant.
\end{abstract}
\maketitle 
The poles of the finite temperature gluon propagator 
can be related to screening of chromoelectric 
and chromomagnetic fields \cite{gross81}.
The static chromoelectric (Debye) screening mass was
calculated in leading order of perturbation theory long ago
and was found to be gauge independent to this order \cite{gross81}.
The existence of a static chromomagnetic screening mass
generated non-perturbatively  
was postulated by Linde to render the perturbative expansion of
different thermodynamic quantities finite \cite{linde80}.
Beyond the leading order also the Debye screening mass is 
non-perturbative \cite{rebhan94}, i.e depends explicitly on the magnetic mass.
Static electric and magnetic propagators and the corresponding
screening masses were studied in SU(2) gauge theory in Landau gauge
\cite{heller95}. In Ref. [5] it was shown
that static propagators could be studied in the effective dimensionally
reduced version of finite temperature SU(2) gauge theory, the 3d adjoint
Higgs model. Here we will discuss the
determination of the screening masses from  propagators
calculated in different gauges and their gauge dependence. 
Results calculated within  full 4d SU(2) gauge theory as well as  the effective
3d theory will be presented.

In four dimensions (4d) all our calculations are performed with the standard
Wilson action for SU(2) lattice gauge theory.
We will use the notation
$\beta_4={4 /g_4^2}$ for the lattice gauge coupling. In three
dimensions  the standard dimensional reduction process leads us to 
consider the 3d adjoint Higgs model
\ba
&
S=-\beta_3 \sum_P \half \tr U_P 
\nonumber\\
&
-\beta_3 \sum_{\bx,\hat \mu} \half \tr
A_0(\bx) U_\mu(\bx) A_0(\bx+\hat \mu)
U_\mu^{\dagger}(\bx)+\nonumber\\
&
{1\over 4}\beta_3 \sum_{\bx} \biggl[\left(6+h\right) \tr A_0^2(\bx)+ 
x { \left( \tr
A_0^2(\bx)\right)}^2 \biggr],
\label{act}
\ea
where $\beta_3$ now is related to the dimensionfull $3d$ gauge coupling and the 
lattice spacing $a$, {\it i.e.}  $\beta_3={4 / g_3^2a}$,  and the adjoint Higgs 
field is parameterized 
by hermitian matrices $A_0=\sum_a \sigma^a
A_0^a$ ($\sigma^a$ {are the usual Pauli matrices}).
Furthermore, $x$ parameterizes the quartic self coupling of the Higgs field 
and $h$ denotes the bare Higgs mass squared. 
We also note that the indices $\mu,~\nu$ run from 0 to 3 in  four
dimensions and from $1$ to $3$ in three dimensions. 

As we want to analyze properties of the gluon propagator, which is a gauge dependent 
quantity, we have to fix a gauge on each configuration on which we want to calculate
this observable. In the past most studies of the gluon propagator have been performed
in Landau gauge. Here we will consider a class of $\lambda$-gauges  
introduced in
Ref. [6].  
In the continuum these gauges 
correspond to the gauge condition
\be
\lambda \d_0 A_0+\d_i A_i=0.
\ee
Here the index $i$ runs from 1 to 3.
The case $\lambda=1$ corresponds to the usual Landau gauge.
In addition to the $\lambda$-gauges we also consider the Maximally Abelian gauge
(MAG) \cite{kronfeld87}.
In this case one has to fix a residual gauge degree of freedom which we do by
imposing an additional U(1)-Landau gauge condition \cite{amemiya99}. 
In the 4d SU(2) gauge theory we also consider the static time averaged Landau gauge
(STALG) introduced in Ref.[9]. In continuum it is defined by 
\be
\d_0 A_0(x_0,\bx)=0,~~~~ \sum_{x_0} \sum_{i=1}^3 \d_{i} A_{i}=0.
\ee

While the notion of Landau and Maximally Abelian gauges carries 
over easily to the 3d case
we have to specify our notion of $\lambda$-gauges in $3d$. We have 
considered two versions of $\lambda$-gauge,
\be
\d_1  A_1 + \d_2  A_2 +\lambda_3  \d_3  A_3=0,
\ee
\be
\lambda_1 \d_1 A_1 + \d_2  A_2 +\d_3  A_3=0,
\ee
which we will refer to as $\lambda_1$-gauge and $\lambda_3$-gauge
correspondingly.
Furthermore, we consider in 3d the
Coulomb gauge,
which fixes the gauge in a plane transverse to the $z$-direction,
$\d_1 A_1 + \d_2  A_2  =0$.
The residual gauge freedom is fixed by demanding that 
$\sum_{x,y} U_3(\bx)=u_{30}$ should be  constant.

All our 4d simulations have been performed at temperature $T=2 T_c$.
The values for 
the 4d coupling
$\beta_4$ corresponding to this temperature were taken from Ref. [4],
$\beta_4=2.52$ for $N_t=4$ and $\beta_4=2.74$ for $N_t=8$.
The 3d  adjoint Higgs model  simulations were done for three sets of parameters,
$\beta_3=11$, $x=0.099$, $h=-0.395$ corresponding to $T=2 T_c$, and 
$\beta_3=16$, $x=0.03$, $h=-0.2085$ as well as $\beta_3=24$, $x=0.03$, $h=-0.1510$ 
which correspond to $T \sim 9200T_c$. For the 3d gauge coupling $g_3^3$ we
always use the 1-loop relation $g_3^2=g^2(T) T$ with $g(T)$ being the 
4d 1-loop running coupling constant in $\overline{MS}$-scheme with 
$\bar \mu=18.86 T$
\cite{long}. The detailed procedure for fixing the parameters
of the 3d effective theory is described in Ref. [12].

We note that for static fields the gauge condition
for 4d $\lambda$ -gauges, $\lambda \d_0 A_0+ \d_i A_i=0$ 
is equivalent to 3d Landau gauge. The
same is true for STALG. One would expect that if dimensional reduction works,
propagators calculated in different $\lambda$ -gauges and STALG  agree
with each other, and of course, with the  propagators calculated in the 3d 
effective theory in Landau gauge. 
A comparison of the corresponding data in 4d SU(2) gauge theory
and the 3d effective theory shows that this is indeed the case. 
Furthermore,
we have compared the electric and magnetic propagators calculated in 4d SU(2)
gauge theory and 3d effective theory in MAG. Good agreement between 3d and 4d
data was found also here.

\begin{figure}
\centerline{a}
\epsfxsize=4cm
\epsfysize=4cm
\centerline{\epsffile{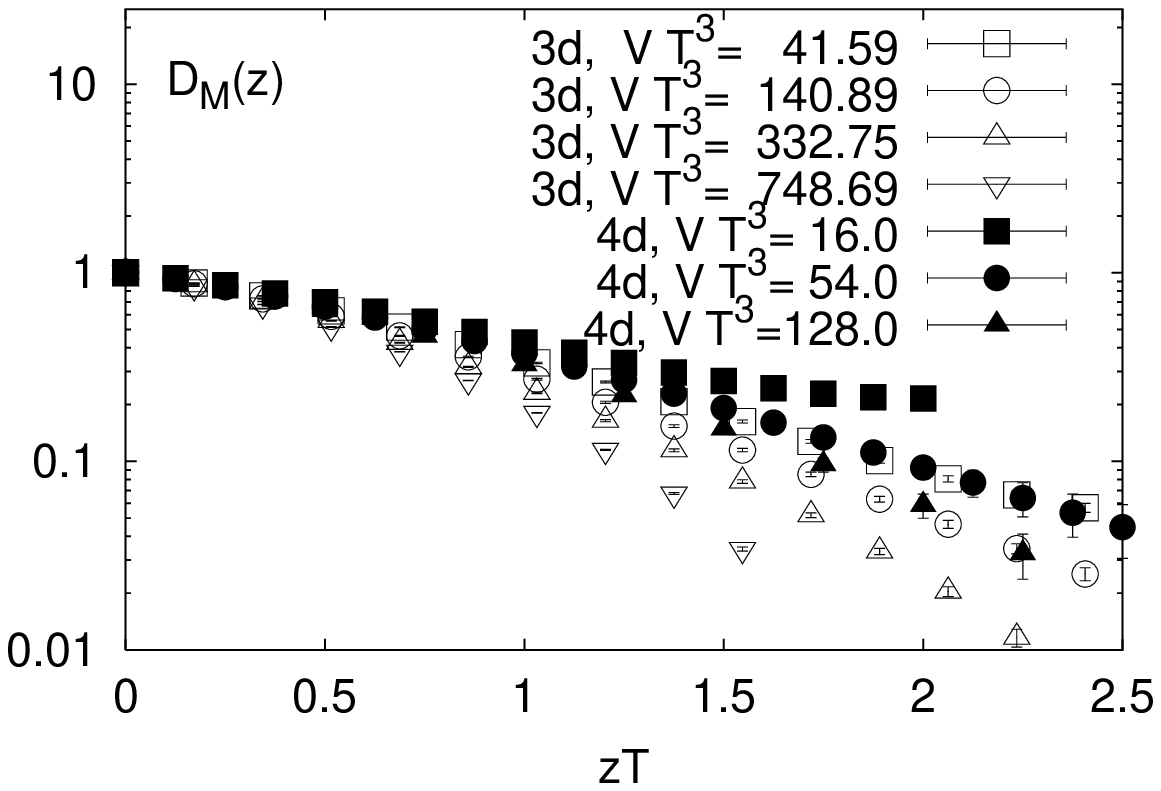}}
\centerline{b}
\epsfxsize=4cm
\epsfysize=4cm
\centerline{\epsffile{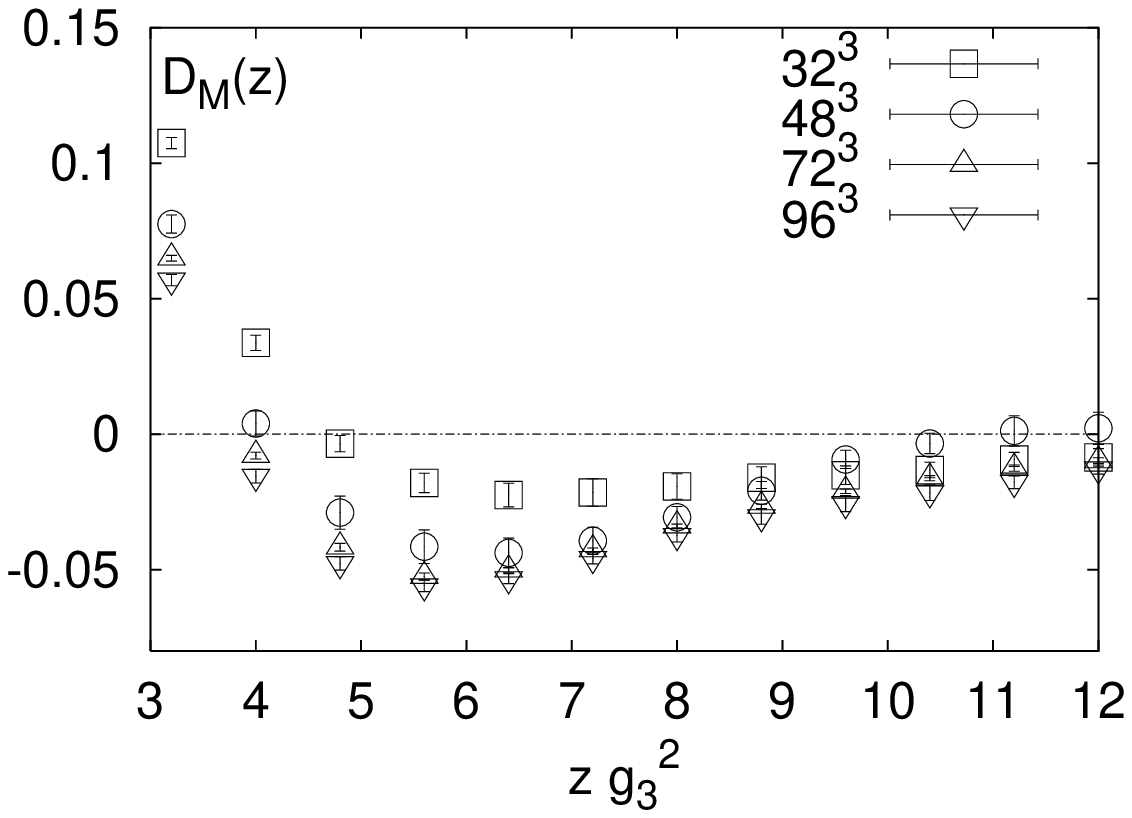}}
\centerline{c}
\epsfxsize=4cm
\epsfysize=4cm
\centerline{\epsffile{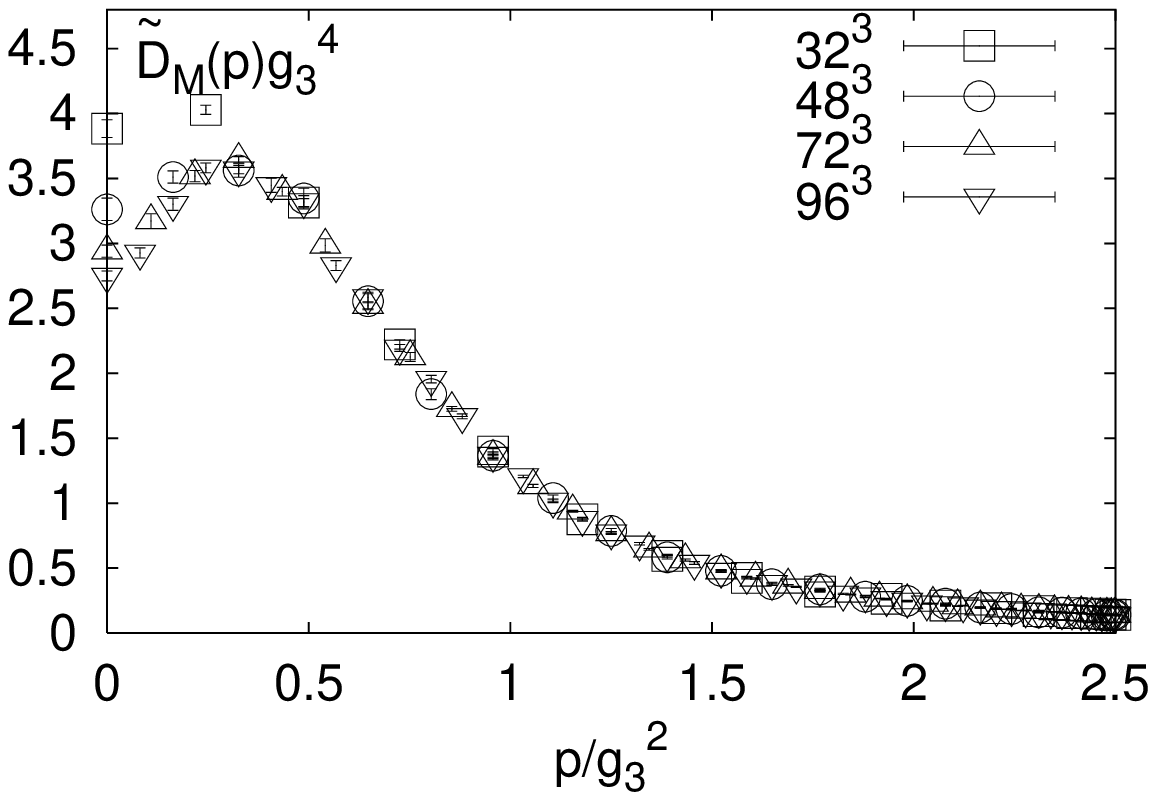}}
\vspace*{-0.5cm}
\caption{
The magnetic gluon propagators in Landau gauge.
Shown are the magnetic propagators in coordinate space
from 4d simulations and from 3d pure gauge theory at $\beta_3=8$ (a),
the magnetic propagators in coordinate space at large distances calculated
in 3d pure gauge theory at $\beta_3=5$ (b) and the momentum space magnetic
propagators from 3d pure gauge theory at $\beta_3=5$ (c). 
The coordinate space propagators were normalized to 1 at $z=0$.
}
\label{magn}
\vspace*{-1.0cm}
\end{figure}

Previous lattice calculations of the magnetic propagator in hot SU(2) gauge theory in
4d \cite{heller95} and in the 3d adjoint Higgs model \cite{karsch98}
gave evidence for its exponential decay
in coordinate space and thus suggested the existence of a magnetic mass.  
In was also found that the magnetic propagators calculated in the 3d adjoint Higgs model
are quite insensitive to the scalar couplings and are quite close to
the corresponding propagators of 3d pure gauge theory.
In Ref. [11]
the Landau gauge gluon propagator of 3d gauge theory was studied in momentum
space and was found to be infrared suppressed for large volumes. Such a behavior
clearly rules out the existence of a simple pole mass. 
To clarify the picture of magnetic screening a detailed study of finite size
effects is necessary.
In what follows we will mainly concentrate on a discussion 
of the magnetic propagators in the limit
of the 3d pure gauge theory. Where appropriate, comparison with the 
results from
4d SU(2) gauge theory will be made.
In Figure \ref{magn}a we show the magnetic propagators
in coordinate space on different volumes at $T=2 T_c$.
Calculations were done in 4d SU(2) gauge theory at $\beta_4=2.52,~2.74$
and in 3d pure gauge theory at $\beta_3=8$.  
On small volumes the propagator indeed shows exponential decay
but it continues to drop faster as the volume increases.
For volumes  $VT^3 \gsim 330 $ 
we have clear evidence that
the propagators become negative at $zT \gsim 2$.
A similar strong volume dependence was observed in other
$\lambda$-gauges and also in Coulomb gauge.

Because of the strong volume dependence of the magnetic propagators in
$\lambda$-gauges simulations on large lattices are necessary
to get control over finite size effects.
We have analyzed
the magnetic propagators in Landau gauge in the 3d pure gauge theory
for $\beta_3=5$ and on lattices of size $L^3$ with $L=32,40,48,56,64,72$ and
$96$.  
In what follows all dimensionfull quantities will be scaled by appropriate
powers of the 3d gauge coupling $g_3$. Using the relation of $g_3$ to the
renormalized 4d gauge coupling (see above) 
it is  straightforward to express dimensionfull
quantities in units of $T$.
The magnetic propagators at $\beta_3=5$ and for
different volumes are shown in Figure \ref{magn}b and Figure 1c.
As one can see from the figure the volume dependence of the coordinate space
propagator is small on these large lattices. The propagator becomes
negative for $z g_3^2 \gsim 4$.  
The strong volume dependence translates into infrared sensivity of the momentum
space propagators.
For $p/g_3^2<0.3$ the  momentum space propagator is
sensitive to the volume, while for large momenta ($p/g_3^2 > 0.3$) it is essentially
independent of the lattice size.
Moreover, we note that for small momenta the finite volume effects lead
to a decrease of $\tilde{D}(p)$ with increasing volume while the volume
dependence is already negligible for $p/g_3^2 \simeq 0.3$. Thus the magnetic
propagators in Landau gauge are infrared suppressed. The propagators in
other $\lambda$-gauges show similar behavior. Let us note that the infrared
suppression of the Landau gauge propagator was observed also $T=0$ 4d SU(N)
gauge theory \cite{alkofer}.

In contrast to the complicated structure found in Landau gauge
the propagator calculated in MAG does show a simple exponential decay at
large distances and does not show any significant volume dependence. 
We find for the magnetic screening mass
in MAG $m_M=0.50(5)g_3^2$.

\begin{figure}
\epsfxsize=4.5cm
\epsfysize=4.5cm
\centerline{\epsffile{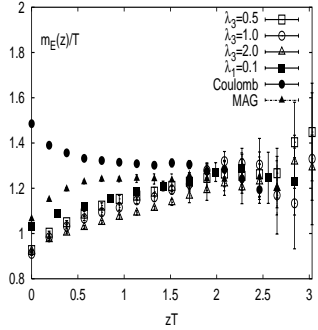}}
\vspace*{-0.5cm}
\caption{Local electric masses calculated in $\lambda$-gauges,
Coulomb gauge and MAG at $T \sim 9200 T_c$. Calculation were done 
in the 3d effective theory at $\beta_3=24$, $x=0.03$ and $h=-0.1510$
except in the case of $\lambda_1=0.1$ where they were done
at $\beta_3=16$, $x=0.03$, $h=-0.2085$.
}
\label{lme}
\vspace*{-0.7cm}
\end{figure}

Contrary to the magnetic propagators the electric propagators show little
volume dependence and decay exponentially in all gauges considered.
We have investigated in detail the gauge dependence of the
electric propagators in the 3d effective theory at temperature
$T \sim 9200 T_c$. The results are summarized in Figure \ref{lme}.
As one can see from this figure any possible gauge dependence of the local electric
masses in the region where they reach a plateau is statistically
insignificant.
\vskip0.2truecm
\noindent
{\bf Acknowledgments:}
\noindent
The work has been supported by the TMR network
ERBFMRX-CT-970122 and by the DFG under grant Ka 1198/4-1.
The numerical calculations have partly been performed at the HLRS in
Stuttgart and the $(PC)^2$ in Paderborn.

\end{document}